\documentclass[submitting]{nst}

\usepackage{subfigure,dcolumn}
\usepackage[T2A,T1]{fontenc}
\usepackage[russian,english]{babel}

\usepackage{listings}
\begin{document}
\title{Afterpulse measurement of JUNO 20-inch PMTs}
\thanks{This work was supported by Strategic Priority Research Program of the Chinese Academy of Sciences (XDA10011100), the Joint Institute of Nuclear Research (JINR), Russia and Lomonosov Moscow State University in Russia, the joint Russian Science Foundation (RSF),  DFG (Deutsche Forschungsgemeinschaft) , and National Natural Science Foundation of China (NSFC).
The authors acknowledge all colleagues from JUNO collaboration 
for operating the 20-inch PMT testing system.
}
\author{Rong Zhao}
\affiliation{School of	 Physics, Sun Yat-sen University, Guangzhou 510275, China}
\author{Nikolay Anfimov}
\affiliation{Joint Institute for Nuclear Research, Dubna 141980, Russia}
\author{Yu Chen}
\email[Corresponding author, School of	 Physics, Sun Yat-sen University, Guangzhou, 510275, China, Email: ]{cheny73@mail.sysu.edu.cn
}
\affiliation{School of	 Physics, Sun Yat-sen University, Guangzhou 510275, China}
\author{Hang Hu}
\affiliation{School of Physics, Sun Yat-sen University, Guangzhou 510275, China}
\author{Jun Hu}
\affiliation{Institute of High Energy Physics, Chinese Academy of Sciences, Beijing 100049, China}
\author{Xiaolu Ji}
\affiliation{Institute of High Energy Physics, Chinese Academy of Sciences, Beijing 100049, China}
\author{Denis Korablev}
\affiliation{Joint Institute for Nuclear Research, Dubna 141980, Russia}
\author{Min Li}
\affiliation{Institute of High Energy Physics, Chinese Academy of Sciences, Beijing 100049, China}
\author{Alexander Olshevskiy}
\affiliation{Joint Institute for Nuclear Research, Dubna 141980, Russia}
\author{Zhaoyuan Peng}
\affiliation{Institute of High Energy Physics, Chinese Academy of Sciences, Beijing 100049, China}
\author{Arseny Rybnikov}
\affiliation{Joint Institute for Nuclear Research, Dubna 141980, Russia}
\author{Zhonghua Qin}
\affiliation{Institute of High Energy Physics, Chinese Academy of Sciences, Beijing 100049, China}
\affiliation{State Key Laboratory of Particle Detection and Electronics, Beijing 100049, China}

\author{Jun Wang}
\affiliation{School of	 Physics, Sun Yat-sen University, Guangzhou 510275, China}

\author{Wei Wang}
\email[Corresponding author, School of	 Physics, Sun Yat-sen University, Guangzhou, 510275, China, TEL: 020-84113269, Email: ]{wangw223@mail.sysu.edu.cn
}
\affiliation{School of	 Physics, Sun Yat-sen University, Guangzhou 510275, China}
\affiliation{Sino-French Institute of Nuclear Engineering and Technology, Sun Yat-sen University, Zhuhai 519082, China}

\author{Zhimin Wang}
\affiliation{Institute of High Energy Physics, Chinese Academy of Sciences, Beijing 100049, China}
\author{Bj\"orn Wonsak}
\affiliation{Institut f\"ur Experimentalphysik, Universit\"at Hamburg, Hamburg 20148, Germany}

\begin{abstract}
 In this article we present the large photo-multiplier tube (PMT) afterpulse measurement results of Jiangmen Underground Neutrino Observatory (JUNO) experiment. Totally 11 dynode-PMTs (R12860) from Hamamatsu company\footnote{Hamamatsu Photonics K.K. (HPK)
} and 150 
 micro-channel plate PMTs (MCP-PMTs, GDB-6201) from NNVT company\footnote{North Night Vision Technology Co., Ltd. (NNVT)
}were tested, an afterpulse model is built according to the afterpulse time distribution and probability of occurrence for these two types of PMTs. 
The average ratio between the total afterpulse charge with the delay between 0.5$~\mu$s and 20$~\mu$s to the primary pulse charge is $\sim$5.6\% (13.2\%) for the tested MCP-PMTs (dynode-PMTs). 
JUNO experiment will deploy 20,012 20-inch PMTs, and this study will benefit the detector simulation,  event reconstruction and data analysis of JUNO experiment.

\end{abstract}

\keywords{Afterpulse; MCP-PMT; JUNO; dynode-PMT; 20-inch PMT }

\maketitle

\section{Introduction}

Jiangmen Underground Neutrino Observatory (JUNO) is a reactor anti-neutrino experiment 
designed 
with multiple physics goals, including determining the mass
ordering of the neutrinos and high-precision measurement of neutrino oscillation parameters~\cite{JUNO:2015sjr,An:2015jdp,juno:ppnp}.
 JUNO’s Central Detector (CD) is a Liquid Scintillator detector (LS) with 20 ktons LS enclosed in an acrylic sphere with a diameter~35.4 m.
Ultra-pure water will be filled outside of the acrylic sphere in the water pool. There will be 17,612  20-inch photo-multiplier tubes (PMTs) viewing 
photons generated in CD 
volume, including 5,000 dynode-PMTs
and 12,612 micro-channel plate PMTs (MCP-PMTs)~\cite{JUNO:2015sjr,An:2015jdp}. There are additionally 2,400 20-inch PMTs positioned in the water pool veto system. 
The pulses generated by 20-inch PMTs for one event will be used to reconstruct information about the detected particles.
 
Afterpulses are undesired signals that follow the light-induced photoelectron signal (called primary pulse) of PMTs. Its mechanism has been studied for a long history and is known to have a negative effect on the
timing of PMTs~\cite{coates1973origins,Akchurin:2007mu,ma2011time,Haser:2013is,zhao2016afterpulse,campbell1992afterpulse}. Afterpulses are mainly caused by positive ions from ionization of residual gas inside PMTs~\cite{coates1973origins}, which are hard to be distinguished with the actual light-induced signal. The afterpulse signals will mimic signal and deteriorate the PMT time resolution, thus affecting the particle 
identification and event reconstruction of JUNO experiment. For this consideration, JUNO experiment requires the total afterpulse ratio to be less than 15\% for 20-inch PMTs~\cite{JUNO:2015sjr}.
The 20-inch dynode-PMTs manufactured by HPK use the traditional dynodes as electron magnification system, and there have been many afterpulse testing results of this type of PMTs~\cite{Chang:2016tte,Zhang:2020kkr}. While the afterpulse characters of the newly developed 
20-inch MCP-PMTs are not well understood due to MCP's special electron magnification mechanism~\cite{Chen:2016myy}.
 In this article, we focus on the ion initiated afterpulses which occur later than hundreds of nanoseconds after the primary pulse, while the light-induced late pulses will not be covered~\cite{4326532,Lubsandorzhiev:2006np,pllbp,Brack:2012ig}. The afterpulse measurement is part of JUNO's PMT characterization test from 2017 to 2021, during which period 150 MCP-PMTs and 11 dynode-PMTs were sampled for the afterpulse measurement. The testing data and operation notes were managed by JUNO PMT testing database~\cite{wang2022database}. Section~\ref{exp:setup} shows the experiment setup and waveform data analysis, Section~\ref{sec:aptq} shows the time and charge distribution of afterpulses and a simplified afterpulse model for JUNO PMTs.

\section{Experiment Setup and Afterpulse Testing Method}
\label{exp:setup}
\subsection{ Experiment setup} 
The JUNO PMT instrumentation group built two independent testing systems for performance characterization of 20-inch PMTs: the scanning station which is designed for precise PMT photo-cathode characterization~\cite{Anfimov:2017bsm,performance_2022}, and the container system which is appropriate for PMT mass test~\cite{Wonsak:2021uum,performance_2022}. Both facilities are capable of measuring afterpulse signals of 20-inch PMTs. 
\begin{figure}[!htbp]
    \centering
   \includegraphics[width=.7484\hsize]{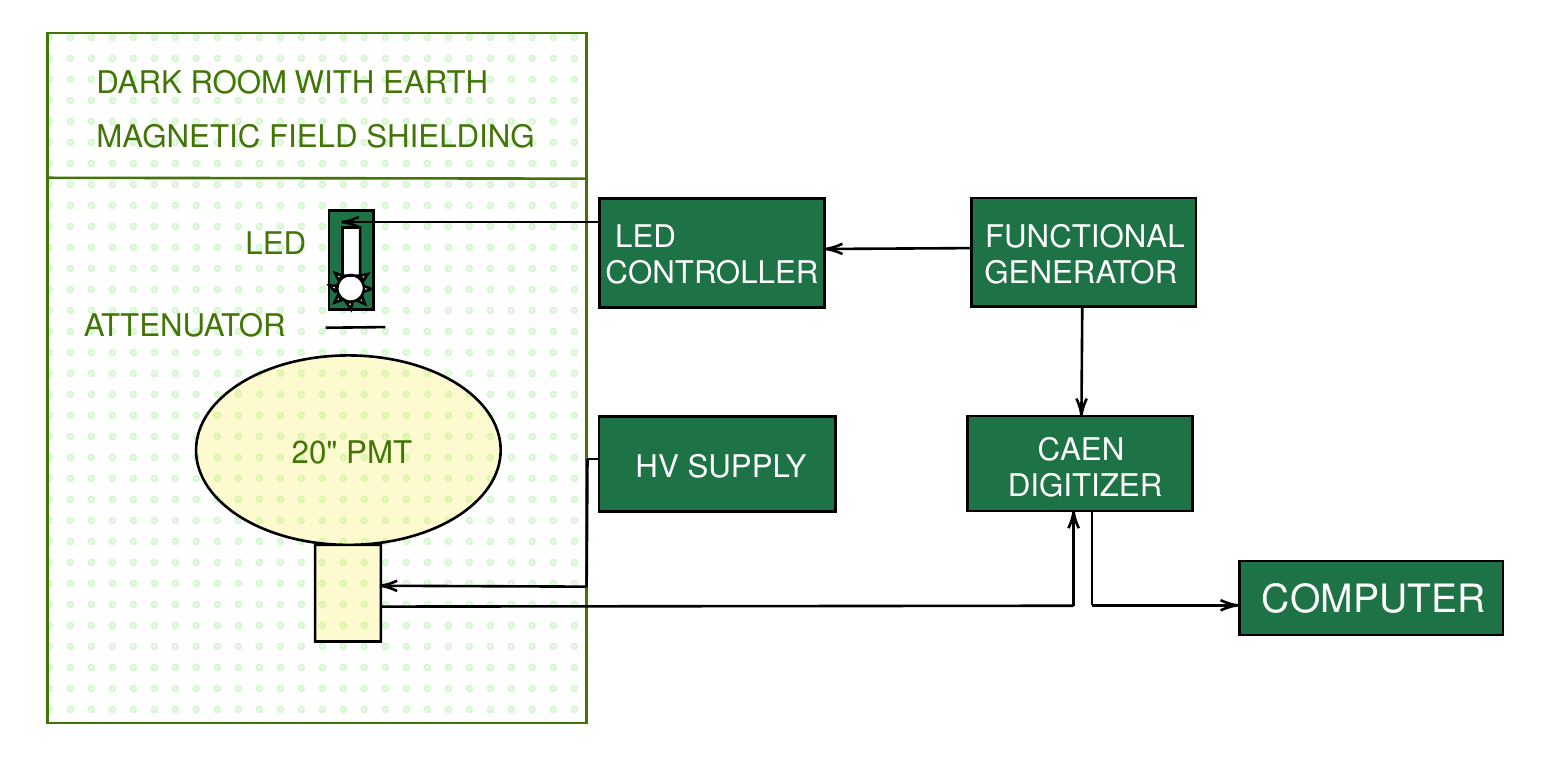}
    \caption{Schematic overview of the testing system in scanning station.}
    \label{fig:syssetup}
\end{figure}
A block diagram of the measurement equipment in the scanning station is shown in Fig.~\ref{fig:syssetup}. During each afterpulse test,
the candidate PMTs will be installed by aluminum holder and connected to the JUNO official PMT base~\cite{JUNO:2015sjr}. The PMT base provides a positive high voltage (HV) to PMT and couples the PMT signal to a waveform digitizer. A 10-bits fast ADC (Analog-to-Digital Converter, CAEN VME1751) is used for waveform digitization. A 420 nm light-emitting diode (LED) is mounted right above the PMT, and an absorptive
ND filter from ThorLab company is mounted between LED and PMT helps to attenuate the light intensity~\cite{tlurl}. The LED has an internal light intensity monitoring and a feedback controlling system to achieve 1\% level intensity stability. 
 The light intensity can be adjusted by LED controller from 0 photoelectrons (p.e.) to hundreds of photoelelectrons per pulse during one afterpulse measurement.
 External trigger signals from
 pulse generator are used to trigger LED and digitizer with 100 Hz frequency. The previous study has shown that large size PMTs are sensitive to earth magnetic field (EMF) due to its influence on the electrons' motion inside PMT~\cite{hamamatsu2007photomultiplier}. To avoid the influence of earth magnetic field, the dark room was designed with earth magnetic field shielding function using three independent groups of Helmholtz coils~\cite{Anfimov:2017bsm}. The residual EMF can achieve a $0.5\ \mu$T level ($\sim 1/10$ of the local EMF intensity) in the centre of the dark room where PMT is mounted.

The container \#4 system adopted JUNO electronics for the waveform digitization~\cite{JUNO:2015sjr}, which can capture a $10\ \mu$s long PMT waveform frame. Instead of
using LED as the light source,
the container electronics work in the self-trigger mode for afterpulse measurement. A relative large trigger threshold ($\sim100 ~$mV) is used so that it can select events with large primary signals. The possible source of those large signals (serve as primary pulse in the afterpulse measurement) in the dark environment could be cosmic rays, radioactive background etc.~\cite{largepulses}. The container afterpulse measurement has independent light source and electronics with scanning station, thus double-checking of the afterpulse timing results. 

\subsection{Afterpulse test method}
It has been confirmed 
that the afterpulse rate and charge depend on the primary pulse intensity~\cite{Akchurin:2007mu,Cheng:2014lka}.
Then for each afterpulse test, we use 3 different light intensities with average intensity $\sim40$ p.e., $\sim80$ p.e., and $\sim120$ p.e. to evaluate the final afterpulse level of PMTs. Also, another test with the same condition but LED turned off is performed to evaluate and thus subtract the contribution from PMT dark count signals. During one afterpulse measurement, the electronics of scanning station record $\sim20,000$ waveform frames ($21~\mu$s period) for each LED light intensity.

\begin{figure}[!htbp]
    \centering
  \includegraphics[width=1.0\hsize]{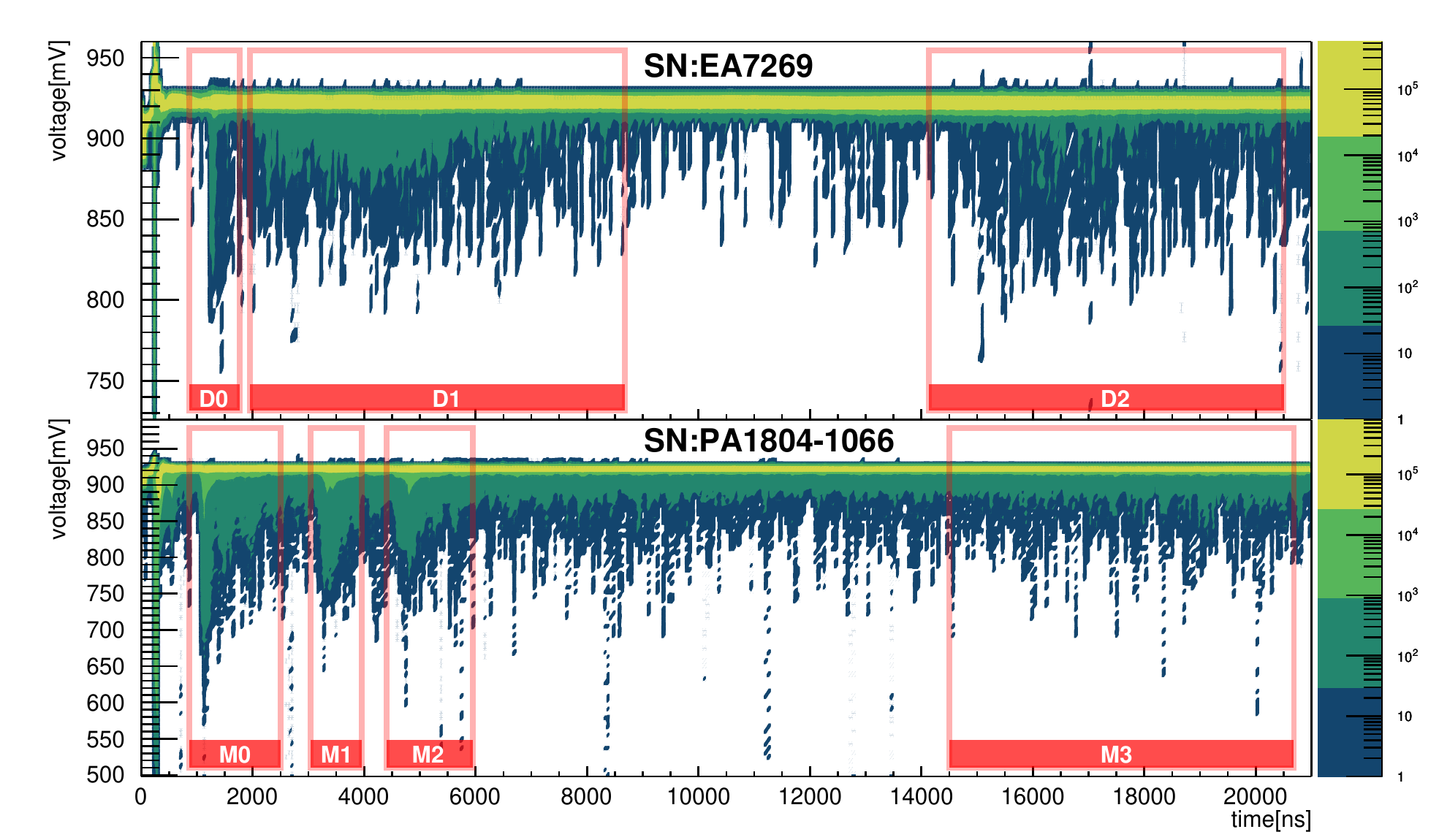}
    \caption{The two-dimensional contour plot by stacking 20,000 tested raw waveform frames. The top panel is from one dynode-PMT with SN "EA7269"; the bottom panel is from one MCP-PMT with SN "PA1804-1066". 
 }
    \label{fig:wave2d}
\end{figure}
As shown in Fig.~\ref{fig:wave2d}, the two-dimensional contour plot by stacking up all the 20,000 waveform frames from one measurement displays the general characteristics of afterpulse signals
such as time and amplitude distribution.
The bottom panel of Fig.~\ref{fig:wave2d} is the afterpulse measurement results from one MCP-PMT with serial number (SN) "PA1804-1066". For this PMT the primary pulse signals arrive at around 300 ns, and three groups of afterpulse signals (arrive at $\sim$ $1.2~\mu$s, $3.7~\mu$s and $5~\mu$s respectively) can be clearly identified. The top panel of Fig.~\ref{fig:wave2d} is the testing results of one dynode-PMT (SN: "EA7269"). The primary pulse signals at $\sim$ 300 ns, the afterpulse signals can be identified at $\sim$ 4.5$~\mu$s and 15$~\mu$s.

The pulses generated by LED light injection are marked as primary pulses;
the pulses arrive 500 ns later than the primary pulse are marked as afterpulses. 
 Some previous afterpulses measurement about 8-inch PMT shown that H$_2^{+}$ induced afterpulses' arrival time is $\sim 200$ ns, and this time delay is the shortest compared to time delay of afterpulses caused by other ions~\cite{ma2011time,Haser:2013is,aiello2018characterisation}. 
 For 20-inch PMTs, the drift time of ions are much longer than 8-inch PMTs. So it is reasonable to address the first group of afterpulses (with time delay around 900 ns) to  H$_2^{+}$ induced afterpulse~\cite{Akchurin:2007mu,ma2011time,Cheng:2014lka,akgun2008afterpulse,Kaether:2012bm,tudyka2016simulation}. The pulses after the primary pulse but with time delay less than 500 ns (later pulses) will be ignored in this study since they are not generated by residual ions. 


\begin{table}[!htbp]
\centering
 \vspace{-.1cm}
\caption{The time windows of different groups of afterpulse signals.}
\label{tab:aptime-table}
\resizebox{\hsize}{!}{%
\begin{tabular}{l|cccc|ccc}
\toprule
PMT Model & \multicolumn{4}{|c|}{MCP-PMT} & \multicolumn{3}{|c}{dynode-PMT} \\ \midrule
afterpulse group  &   M0  &  M1   &  M2   & M3    &  D0     &     D1  &   D2    \\ 
 searching region start [ns]&  500   &2500     & 4000   &15000    & 500      &    1500   &  10000     \\ 
 searching region end [ns]& 2500    & 4000    & 7000    & 21000    &   1500    &  10000     & 21000     \\
  \bottomrule
\end{tabular}%
}
\end{table}
\clearpage

Afterpulses of different population are caused by ions with different time delay, so it is convenient to divide them into groups according to their different time delay. There are four groups of afterpulse confirmed for 20-inch MCP-PMTs, tagged as M0, M1, M2 and M3; and three groups of afterpulse are confirmed for 20-inch dynode-PMTs, tagged as D0, D1 and D2. The time windows for the searching of these afterpulses in data analysis are summarized in Table~\ref{tab:aptime-table}.

The method to classify the afterpulses into groups by their arrival time is not perfect~\cite{ma2011time}, since in one time window there could be multiple sources of afterpulses with different amplitude or charge distribution but share similar time delay~\cite{ma2011time}. In this study, we only focus on the dominant afterpulse signals in one selected time window and ignore the non-significant components which may occur when a much higher HV is applied to PMTs.

\subsection{Afterpulse waveform analysis}

\begin{figure}[!htbp]
    \centering
 \includegraphics[width=\hsize]{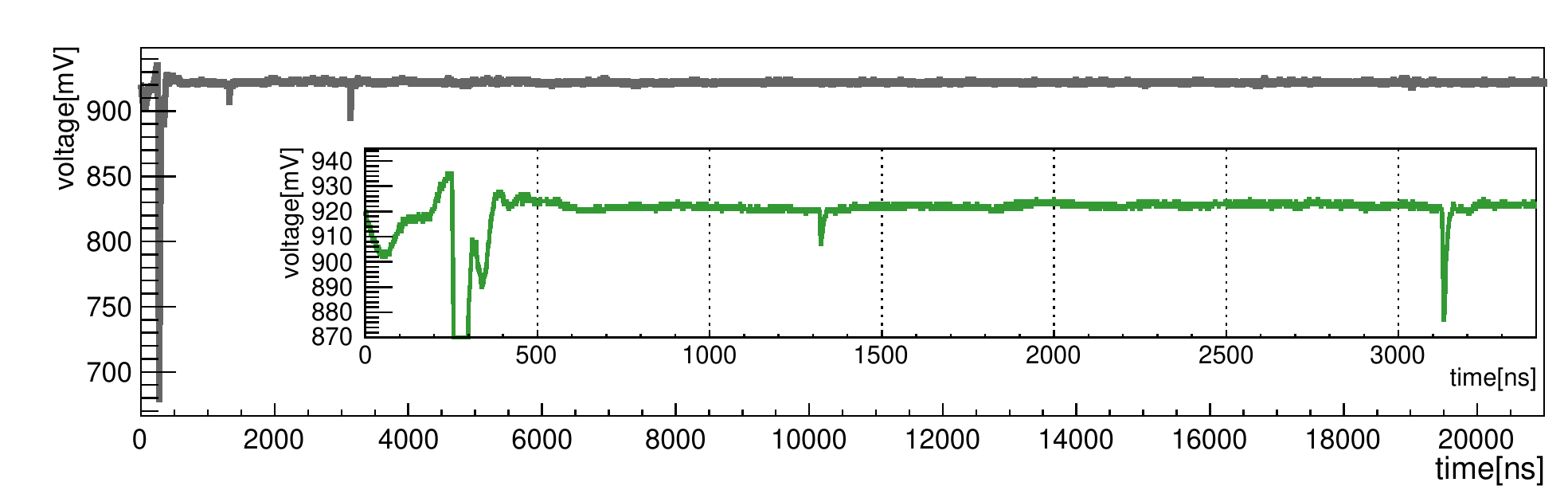} 
    \caption{A tested waveform with 21$~\mu$s length.
    }
    \label{fig:wf1d}
\end{figure}

Figure~\ref{fig:wf1d} is a typical single waveform captured for afterpulse analysis. The primary pulse arrives at around 280 ns in the time axis, two afterpulse signals appear at $\sim 1300$ ns and $\sim3100$ ns.
In the zoom-in pad of Fig.~\ref{fig:wf1d}, it is shown that the baseline around primary pulse is distorted by the pre-pulses, late pulses and potential EM noise~\cite{Lubsandorzhiev:2006np}, so the baseline of primary pulse is calculated as the mean value of waveform from 10$~\mu$s to 20$~\mu$s where the waveform is less affected by LED induced signals. The charge of primary pulse (afterpulse) is integrated from $-25$ ns to $+50$ ns from the peak position of primary pulse (afterpulse). A 3 mV threshold is used to seek afterpulse signals; also the charge of one afterpulse signal is required to be larger than 0.25 single p.e. charge to exclude tiny signals or the low-frequency noise.

For each primary pulse, all the afterpulse signals
in the time range [$T_{\textrm{PP}}$+500, 21000] ns will be recorded for analysis, where $T_{\textrm{PP}}$ is the arrival time of primary pulse. For the baseline calculation of afterpulses, we use the average value over a 50 ns time interval before the threshold-cross time of one afterpulse signal. The arrival time of one afterpulse (primary pulse) signal $T_{\textrm{AP}}$ ($T_{\textrm{PP}}$) is calculated as the threshold-cross time of its leading edge.
 If one afterpulse is found at $T_{i}$, we will search the next afterpulse from $T_{i}+50$ ns, so the minimum time interval between two adjacent afterpulses belonging to one primary pulse is 50 ns in this study.


\section{ Afterpulse Characterization results }\label{sec:aptq}
The timing features and charge distribution of afterpulse signals are two of the most critical parameters for the afterpulse measurement of JUNO 20-inch PMTs. Previous studies have shown that both the charge and rate of afterpulse are dependent on the primary pulse intensity, HV (gain) of PMT and the EMF shielding~\cite{Anfimov:2017bsm}. 
Before each afterpulse test, the working HV of one candidate PMT is tuned to achieve PMT gain\footnote{ MCP-PMT's typical working HV is around 1750V and dynode-PMT's HV is around 1820V.} $G=10^{7}$ and gain is limited to range [$0.95\times10^7$, $1.05\times10^7$].
Then for each PMT test, the EMF shielding is turned on and three different LED light intensities are applied to characterize afterpulses.

\subsection{Timing of afterpulse signals}
The afterpulse time is defined as the time delay between one afterpulse signal and its corresponding primary pulse signal, $ t_{\textrm{AP}}=T_{\textrm{AP}}-T_{\textrm{PP}}$, where $T_{\textrm{AP}}$ ($T_{\textrm{PP}}$) is the arrival time of afterpulse (primary pulse). For one PMT, the afterpulse time of each waveform will be filled into a histogram then fitted by a Gaussian function, the fitted mean value $t^{i}_{\textrm{AP}}$ is then referred as the time for group $i$ ($i=\text{D0}$, D1,...) of afterpulse. 



The afterpulse time distribution histograms and fitting results of one MCP-PMT (SN: "PA1906-2539") are plotted in Fig.~\ref{fig:ap_par}; the afterpulse time distribution histograms and fitting results of one dynode-PMT (SN: "EA7217") are plotted in Fig.~\ref{fig:ap_parham}. For this MCP-PMT, the fitted afterpulse time of M0, M1, M2, and M3 are $t^{\textrm{M0}}_{\textrm{AP}}=906\pm 1 ~\text{ns}$, $t^{\textrm{M1}}_{\textrm{AP}}=3209\pm 3~\text{ns}$, $t^{\textrm{M2}}_{\textrm{AP}}=4669\pm 2~\text{ns}$ and $t^{\textrm{M3}}_{\textrm{AP}}=16.57\pm 0.13~\mu\textrm{s}$, respectively. For this dynode-PMT, the fitted afterpulse time of D0, D1 and D2 are $t^{\textrm{D0}}_{\textrm{AP}}=837\pm 1~\text{ns}$, $t^{\textrm{D1}}_{\textrm{AP}}=3651\pm 14~\text{ns}$, and $t^{\textrm{D2}}_{\textrm{AP}}=13990\pm 14~\text{ns}$, respectively.
The uncertainty of $t^{i}_{\textrm{AP}}$ is estimated considering both the contribution from primary pulse and afterpulse.

\begin{widetext}
\begin{figure*}[htbp]
    \centering
   \includegraphics[width= \hsize]{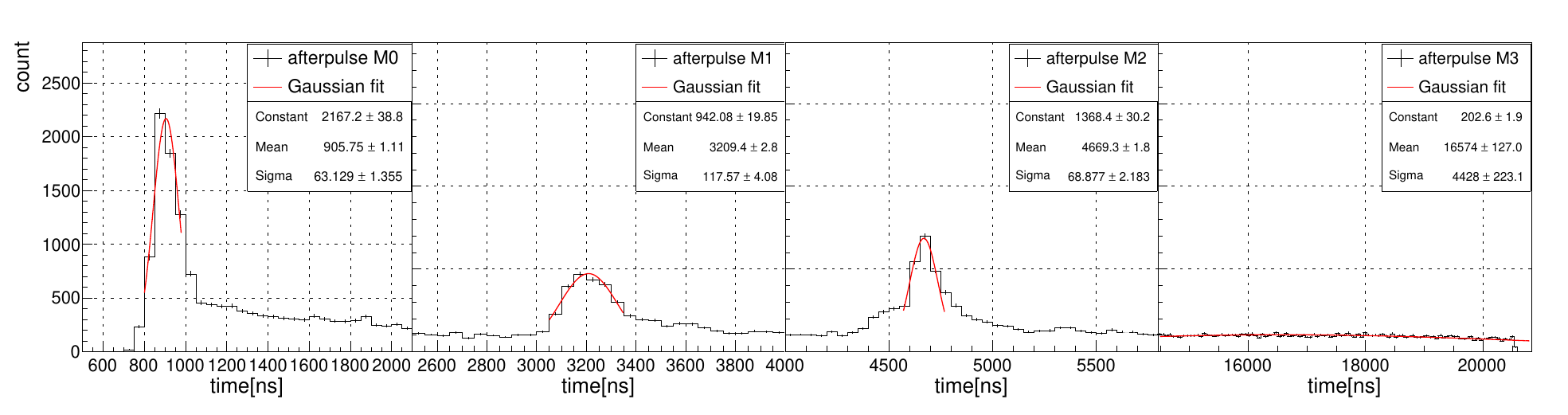}
   \vspace{-.5cm}
    \caption{Gaussian fitting of afterpulse time distribution for one MCP-PMT, SN: "PA1906-2539".   }
    \label{fig:ap_par}
\end{figure*}
\vspace{-1cm}
\begin{figure*}[htbp]
    \centering
   \includegraphics[width=\hsize]{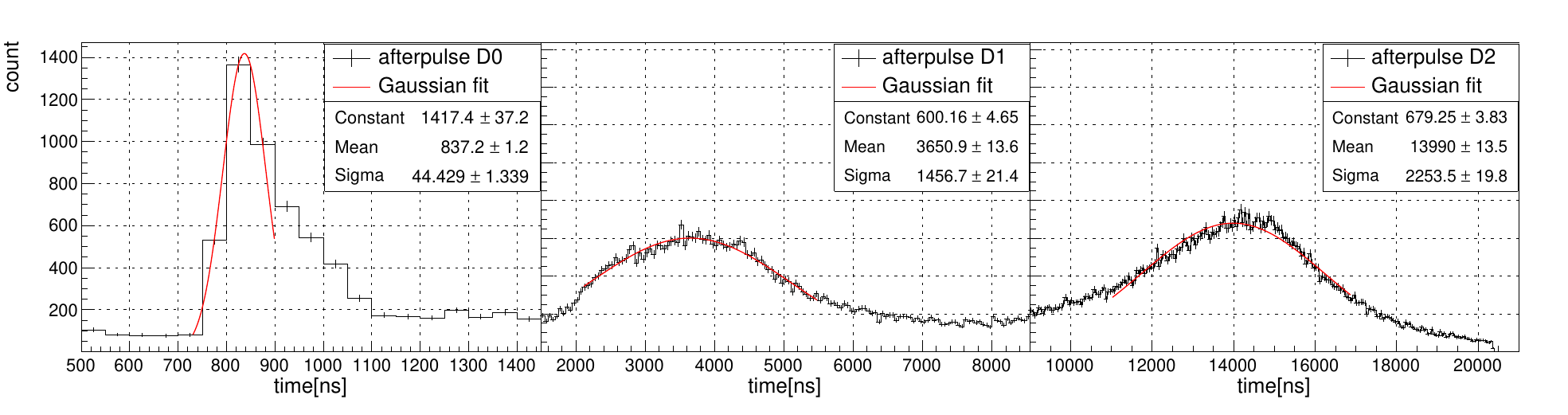}
   \vspace{-.5cm}
    \caption{Gaussian fitting of afterpulse time distribution for one dynode-PMT, SN: "EA7217".  }
    \label{fig:ap_parham}
\end{figure*}
\end{widetext}

\begin{figure}[!htbp]
    \centering
    \includegraphics[width=\hsize]{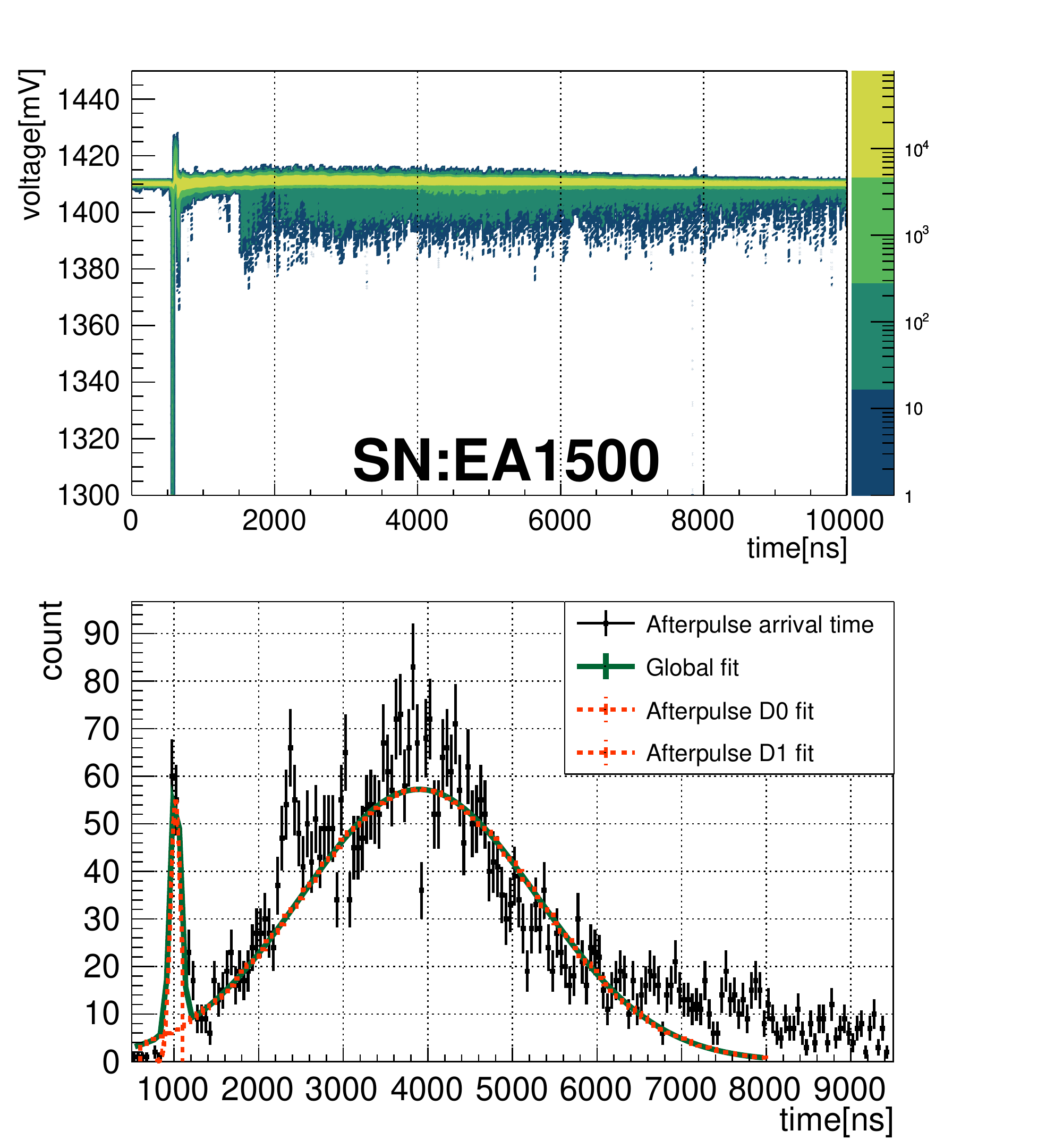}  
    \caption{The two-dimensional  waveform  stacking contour histogram of dynode-PMT "EA1500" (top panel) and the histogram of afterpulse arrival time distribution (bottom panel), measured by container system.}
    \label{fig:charge-timec}
\end{figure}
 \begin{figure}[!htbp]
    \centering
    \includegraphics[width=\hsize]{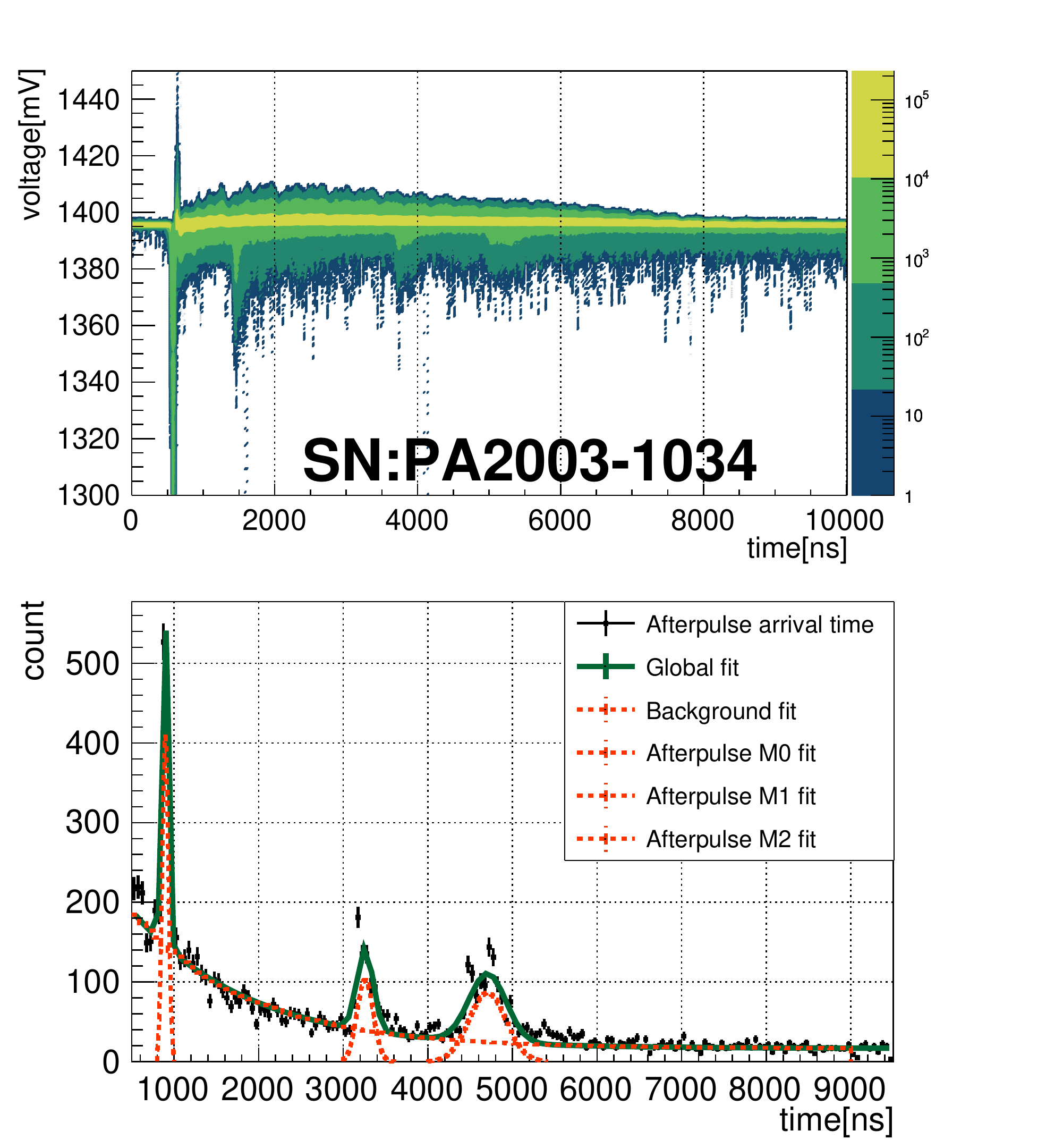}  
    \caption{The two-dimensional  waveform stacking contour histogram of MCP-PMT "PA2003-1034" (top panel) and the histogram of afterpulse arrival time distribution (bottom panel), measured by container system. }
    \label{fig:charge-timemcp}
\end{figure}

 As mentioned in Section~\ref{exp:setup}, the container \#4 PMT test system is also used for afterpulse timing study as a cross check of the scanning station measurement. Due to the limitation of hardware configuration, the electronics of container \#4 system can only record 10$~\mu$s long waveform for afterpulse measurement~\cite{potted}. Fig.~\ref{fig:charge-timec} and Fig.~\ref{fig:charge-timemcp} 
 show the testing results of one MCP-PMT and one dynode-PMT from the container system. The top panel is a two-dimensional stacking contour plot of 20,000 waveform frames and the bottom panel is the histogram of afterpulse time distribution and fitting results. There is some extra noise in the self-trigger mode afterpulse measurement of MCP-PMTs, 
 this kind of signal-induced pulse starts near the end of the primary pulse signal and is exponentially distributed. 

 An exponential background and three Gaussian functions are used to achieve a global fitting of the afterpulse time histogram for MCP-PMT; two Gaussian functions are used for fitting the afterpulse time histogram of dynode-PMT. For MCP-PMT ("PA2003-1034"), the fitted afterpulse time of M0, M1 and M2 are $896\pm 2$ ns, $3259\pm 8$ ns and $4705\pm9$ ns, respectively; for dynode-PMT ("EA1500"), the afterpulse time of D0 (D1) is $1018\pm8$ ns ($3879\pm26$ ns).

 The results from scanning station and container system confirmed the timing features of three groups of MCP-PMT's afterpulse and two groups of dynode-PMT's afterpulse. These two PMT testing systems have independent light source and electronics thus can exclude potential flaws of light source or electronics problems. The time distribution of H$^+_{2}$ induced afterpulse for MCP-PMT is narrower than dynode-PMT; this indicates that the time resolution of MCP-PMT is better than the dynode-PMT. 

The afterpulse timing features of JUNO 20-inch dynode-PMTs are similar to the published results of smaller size dynode-PMT~\cite{Haser:2013is,zhao2016afterpulse}; the afterpulse timing results of MCP-PMTs are consistent with results from vendor~\cite{Wu:2019urv}. 
For each afterpulse component, a weighted average over all the tested PMTs (150 MCP-PMTs and 11 dynode-PMTs ) is calculated as the final afterpulse delay time,
$\overline{t}^{i}_{AP} $ and their uncertainties are shown in Table~\ref{tab:aptime-all}.
\begin{table}[!htbp]
\centering
\caption{The arrival time of each group of afterpulse averaged over all the tested PMTs.}
\label{tab:aptime-all}
\resizebox{\hsize}{!}{%
\begin{tabular}{l|llll|lll}
\toprule
PMT Model & \multicolumn{4}{|c|}{MCP-PMT} & \multicolumn{3}{|c}{dynode-PMT} \\ \midrule
afterpulse group  &   M0  &  M1   &  M2   & M3    &  D0     &     D1  &   D2    \\ 
$\overline{t}^{i}_{\textrm{AP}}$ [ns]&  910 &3134  & 4579   &17731  & 1067     &    4239  &  15081    \\ 
uncertainty of $\overline{t}^{i}_{\textrm{AP}}$ [ns]&  14   &61    & 87   &600   & 132      &    1957   &  2791     \\ 
  \bottomrule
\end{tabular}%
}
\end{table}


\subsection{Charge distribution of afterpulse signals}
Afterpulses with different time delays are caused by different types of ions, each of them has different capability of generating second electrons from photo-cathode. 
\begin{figure}[!htbp]
    \centering
    \includegraphics[width=\hsize]{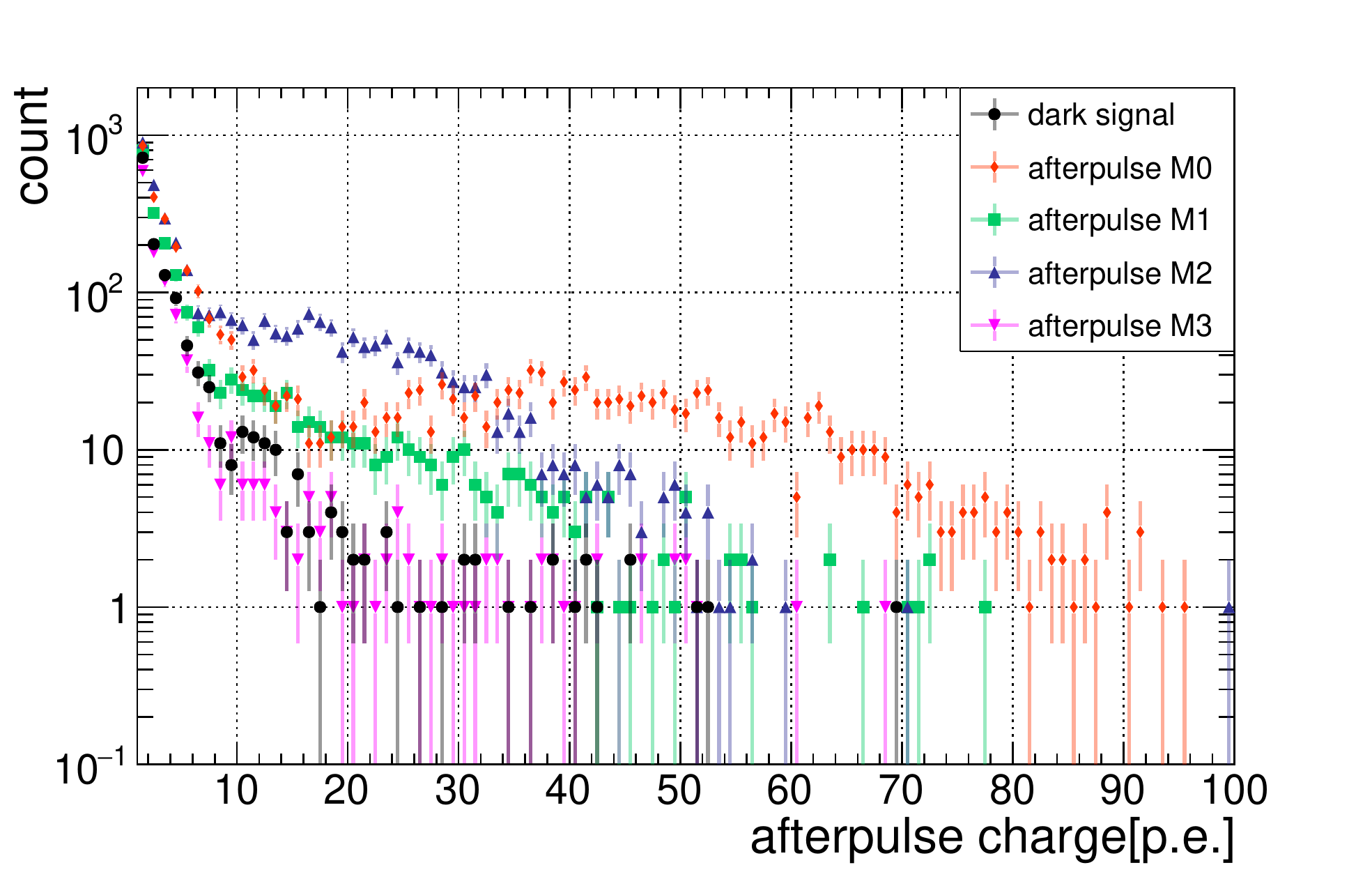}  
        \includegraphics[width=\hsize]{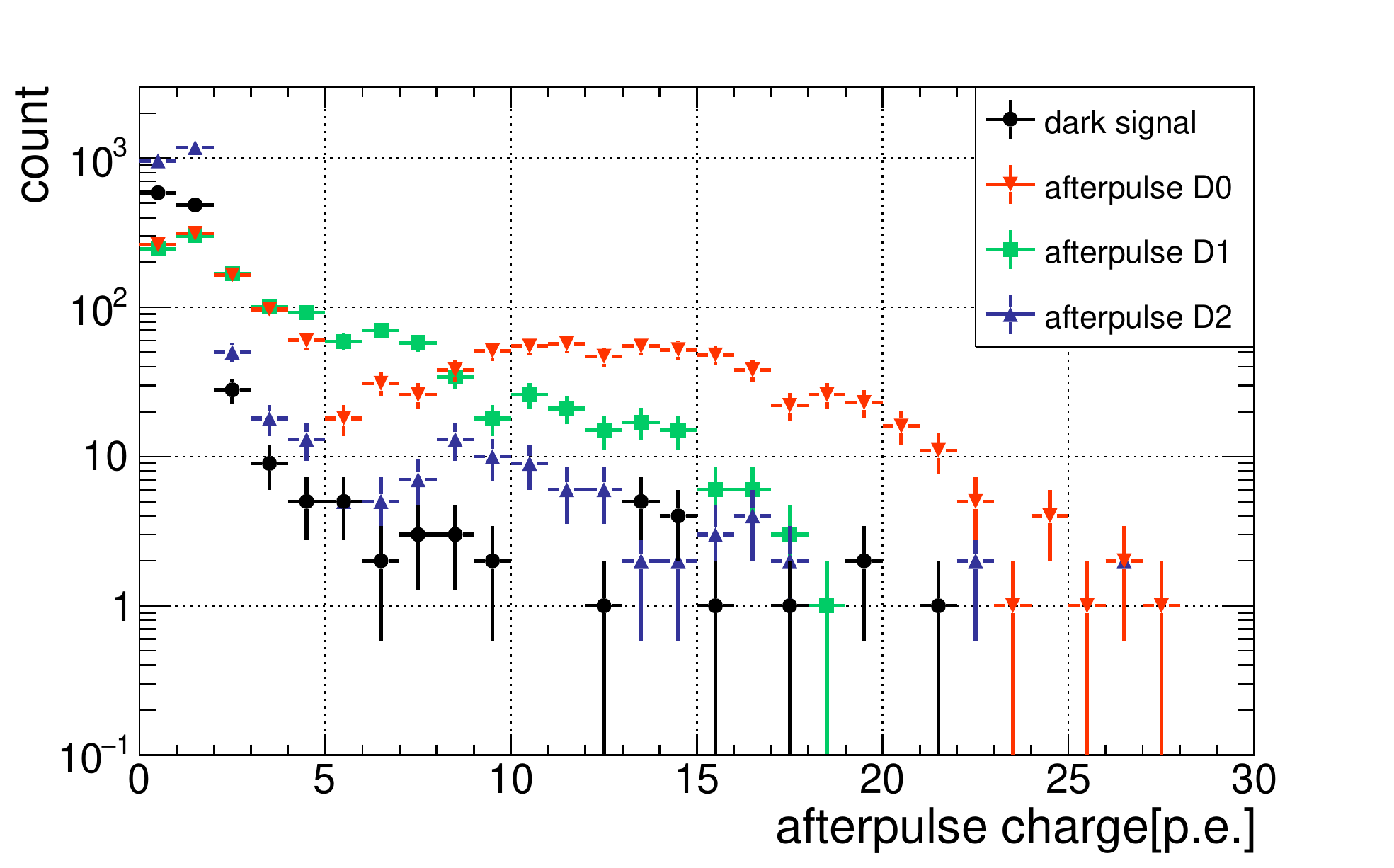}
    \caption{Charge distribution of different groups of afterpulses in number of photoelectrons, the primary pulse intensity is $\sim 120$ photoelectrons. Top panel: MCP-PMT; bottom panel: dynode-PMT.\label{fig:charge_timettnp} }
\end{figure}
Fig.~\ref{fig:charge_timettnp} shows the typical charge distribution of each group of afterpulse when the primary pulse light intensity is $\sim 120$  photoelectrons. The top panel is from one MCP-PMT (SN: "PA1804-1066") and the bottom panel is from one dynode-PMT (SN: "EA7269"). The afterpulse charge of each event is normalized to the primary pulse charge with 100 photoelectrons in the histograms, assuming that the afterpulses' charge is proportional to the primary pulse's charge.
From the charge distribution histograms of MCP-PMT, it is shown that afterpulse M0 has more average photoelectrons than others, which could be the feature of Hydrogen induced afterpulses~\cite{aiello2018characterisation}, H$^{+}_2$ gives a mean secondary yield about 4 electrons and hence quite large afterpulses. The charge of afterpulse M1 and M2 is about tens of photoelectrons level, but most of the M3 afterpulses are single p.e. signal. 
For dynode-PMT, afterpulse D0 also has more average photoelectrons than others; most of D1 and D2 are single photoelectron signals. Compare the results of MCP-PMT and dynode-PMT in Fig.~\ref{fig:charge_timettnp}, afterpulses charge of dynode-PMT are less than 25 photoelectrons when PMT is illuminated by 100 photoelectrons primary pulse while afterpulses of MCP-PMT have more average photoelectrons for each event.  The difference of afterpulse charge distribution between these two types of PMTs is related to  their different electron magnification mechanisms; a detailed study about charge response of MCP-PMT and dynode-PMT can be found in Ref.~\cite{GainandChargeZHQ}. 

\begin{figure}[!htbp]
    \centering
    \includegraphics[width=\hsize]{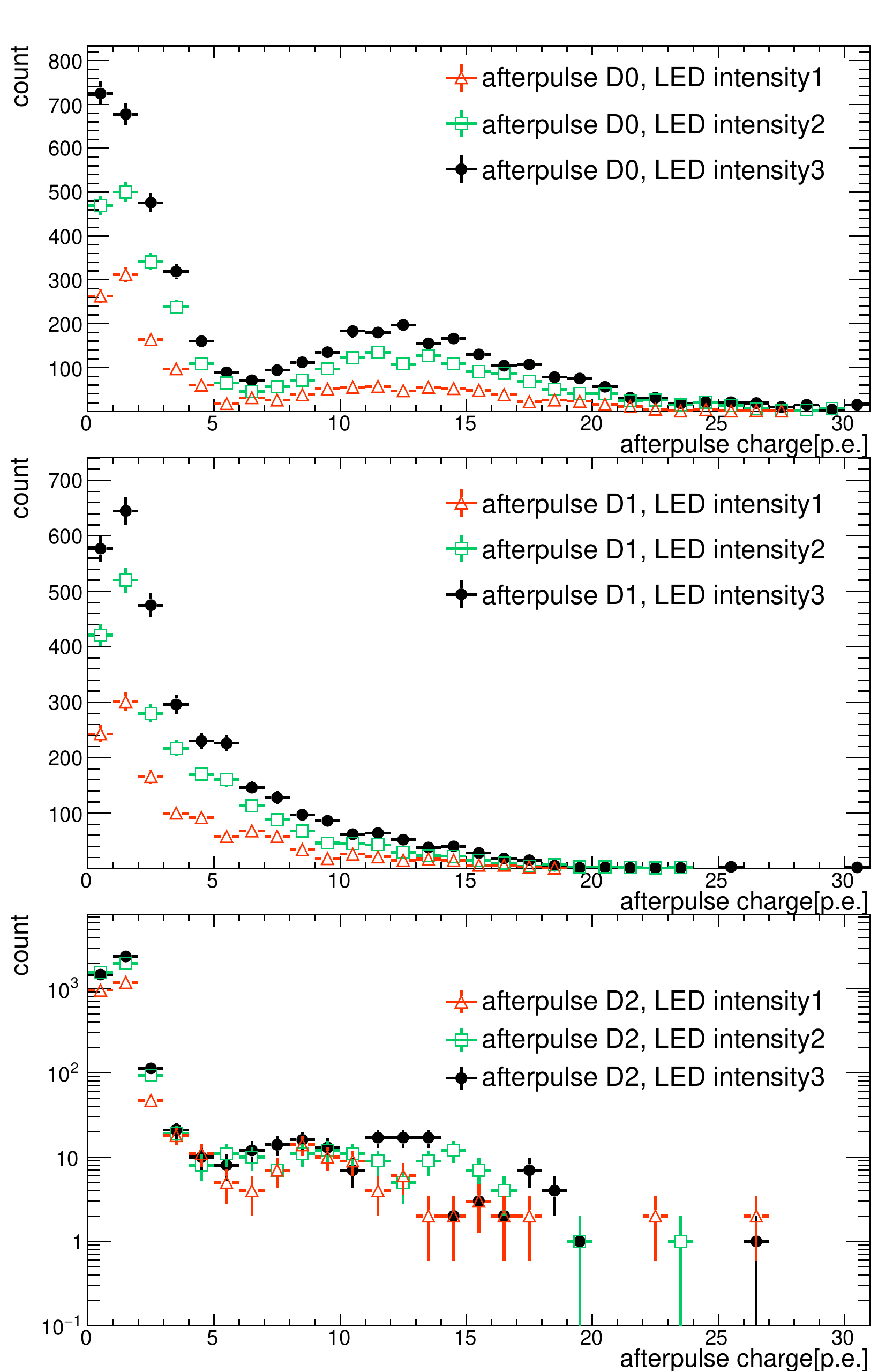}  
    \caption{Charge distribution of different group of afterpulses of one dynode-PMT in number of photoelectrons, the three LED intensities are plotted in different colors. The primary pulse intensity  corresponding to light intensity 1 (2, 3) is $\sim$ 40 (80,120) photoelectrons.}
    \label{fig:charge-lt}
\end{figure}

\begin{figure}[!htbp]
    \centering
    \includegraphics[width=\hsize]{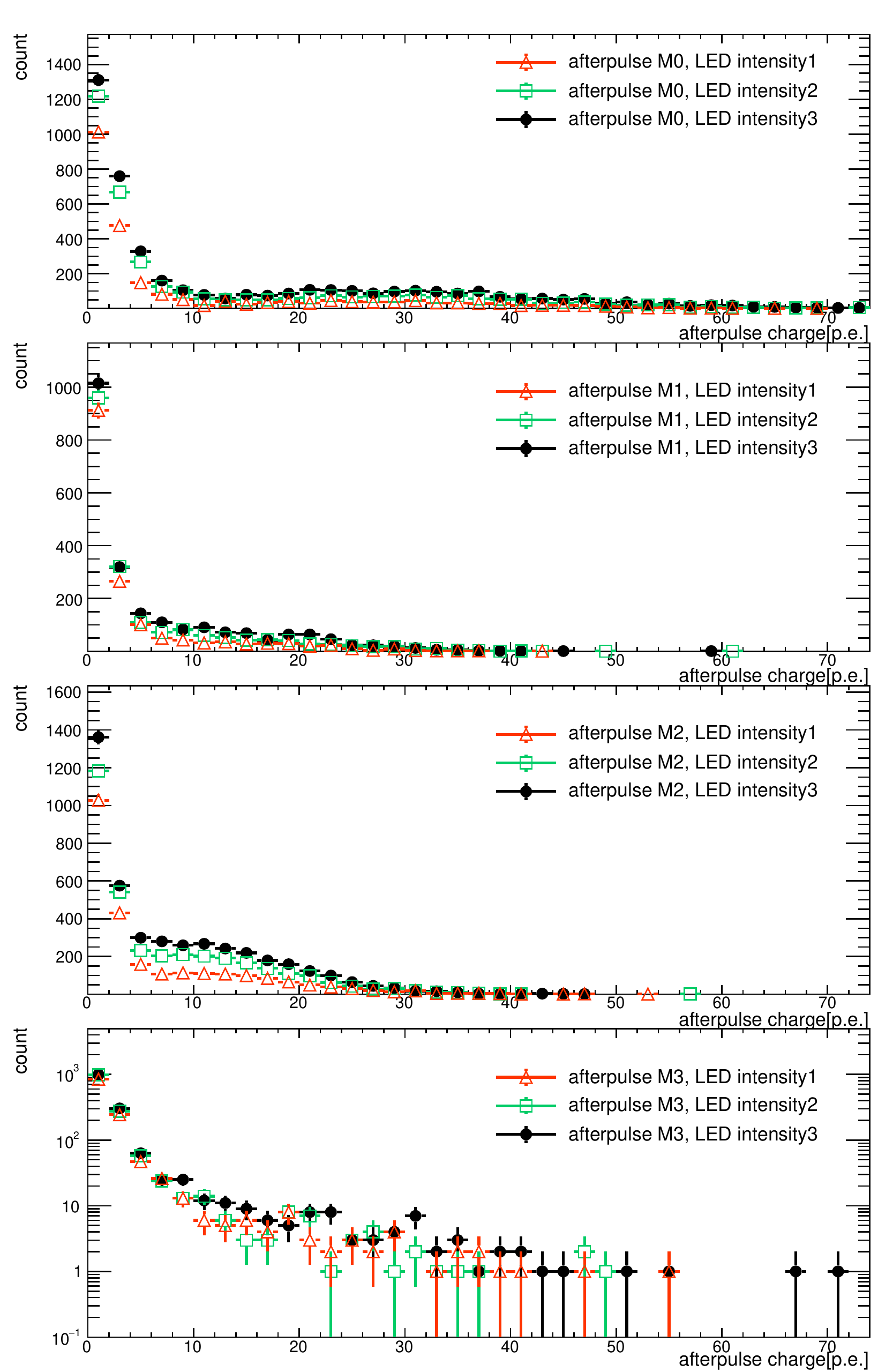}  
    \caption{Charge distribution of different group of afterpulses of one MCP-PMT in number of photoelectrons, the three LED intensities are plotted in different colors. The primary pulse intensity  corresponding to light intensity 1 (2, 3) is $\sim$ 40 (80,120) photoelectrons.}
    \label{fig:mcpcharge-lt}
\end{figure}

The charge distribution of different groups of afterpulse when the PMT is illuminated by different LED intensities are shown in Fig.~\ref{fig:charge-lt} and Fig.~\ref{fig:mcpcharge-lt}. For dynode-PMT, the afterpulse rate of D0 and D1 grow with a higher
primary pulse intensity. The afterpulse rate of D2 keeps stable with primary pulse light intensity. This could be caused by the limited number of heavy residual ions inside the PMT vacuum. 
For MCP-PMT, the rate of afterpulse M0 and M1 also grow with higher primary pulse intensity, while M3 has a similar property to D2, which indates that M3 may also be caused by very heavy ions inside the PMT vacuum.

The afterpulse charge distribution and dependency of afterpulse's rate to primary pulse charge vary for different groups of afterpulse, so it is not appropriate to evaluate afterpulse level by rate. In this case, a total charge ratio between all afterpulses and primary pulse is defined to describe the total afterpulse level for JUNO 20-inch PMTs.
 




\subsection{Afterpulse model of 20-inch JUNO PMTs }

\begin{figure}[!htbp]
    \centering
        \includegraphics[width=\hsize]{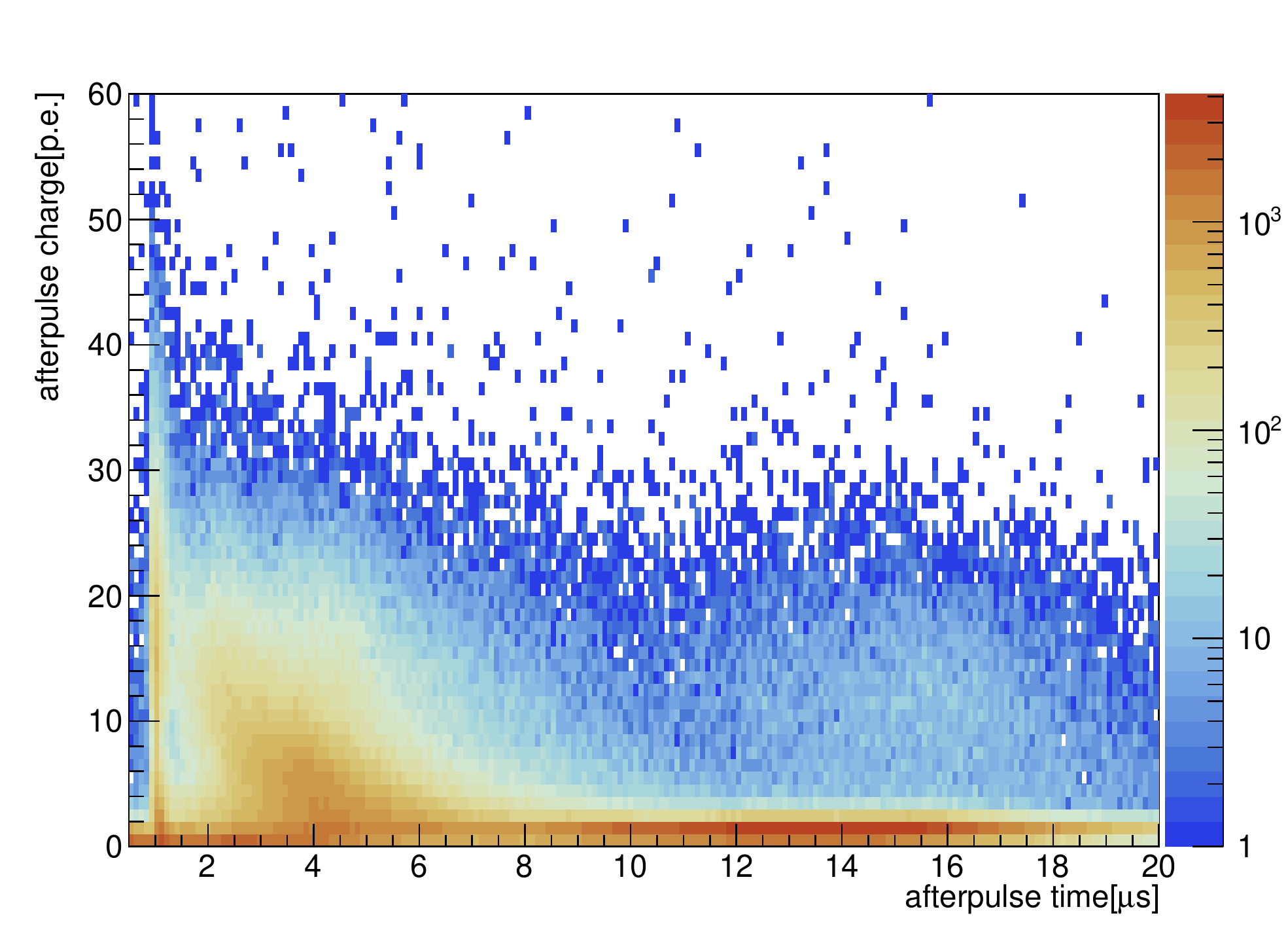}  
        \includegraphics[angle=0,width=\hsize]{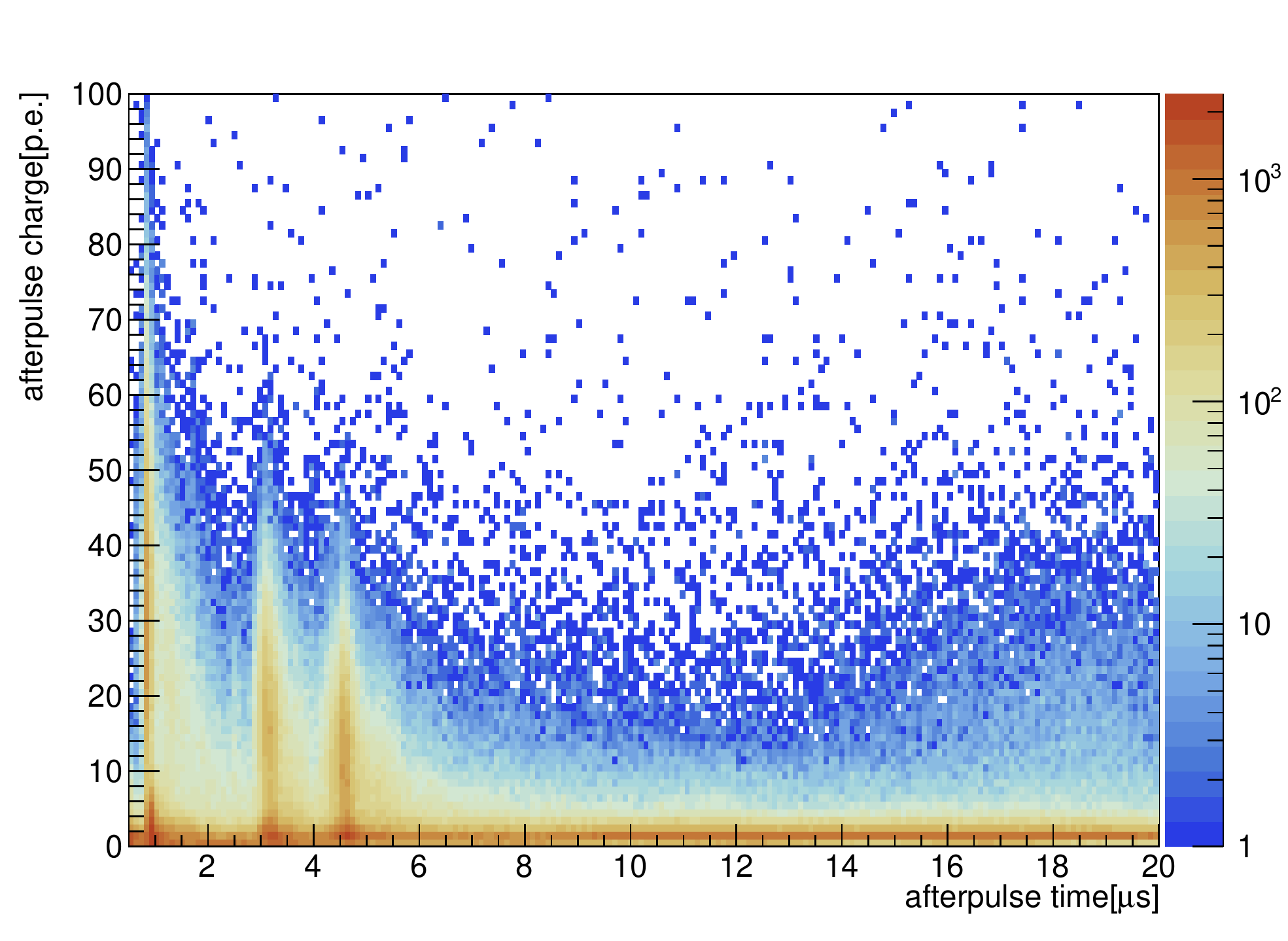}  
    \caption{The two-dimensional  contour plot of the charge-time distribution of afterpulses for dynode-PMTs (top panel) and MCP-PMTs (bottom panel) with light intensity  120 p.e. per trigger.\label{fig:charge-time}}
\end{figure}

The charge and arrival time distribution of afterpulses from all tested PMTs are plotted into a two-dimensional contour plots as shown in Fig.~\ref{fig:charge-time} (with primary pulse $\sim 120$ p.e.).
Since all the PMTs were manufactured with the same technology, the afterpulse timing features are still preserved even after the average process. There are still four (three) groups of afterpulse signals of MCP-PMTs (dynode-PMTs) that can be identified.
For one PMT, the charge ratio of group $i$th of afterpulse $R_{\textrm{AP}}^{i}$ is defined as
\begin{equation}
    R_{\textrm{AP}}^{i}=\frac{\sum_{j} Q_{\textrm{AP}}^{ij}-Q_{\textrm{dark}}^{i}}{\sum_{j} Q_{\textrm{PP}}^{ij}},
\end{equation}
where
$Q_{\textrm{AP}}^{ij}$ is total charge of group $i$ afterpulse signal in frame $j$, and $Q_{\textrm{PP}}^{ij}$ is the charge of primary pulse signal in frame $j$.
$Q_{\textrm{dark}}^{i}$ is total charge contribution from dark signal in the time window of group $i$ afterpulse. 
The total charge ratio $R_{\textrm{AP}}$
of one tested PMT is then defined as
 the sum of all the afterpulse components,
\begin{equation}
    R_{\textrm{AP}}=\sum_{i}R^{i}_{\textrm{AP}}.
\end{equation}
For one PMT, $R_{\textrm{AP}}$ represents the ratio of total charge induced by afterpulses and total charge induced by primary pulses. This value $R_{\textrm{AP}}$ will be treated as the afterpulse occurrence probability of one PMT.
The afterpulse charge ratio of each group of afterpulse and total charge ratio $R_{\textrm{AP}}$ are averaged over all the tested PMTs and summarized in Table~\ref{tab:aptimeratio}. 
\begin{table}[!htbp]

\centering
\caption{The average afterpulse time, charge ratio and rate for each group of afterpulses.}
\resizebox{\hsize}{!}{%
\begin{tabular}{l|llll|lll}
\toprule
PMT Model & \multicolumn{4}{|c|}{MCP-PMT} & \multicolumn{3}{|c}{dynode-PMT} \\ \midrule
 afterpulse group & M0& M1& M2&M3& D0&D1&D2 \\
$R_{\textrm{AP}}$[\%], LED intensity 1&2.23  &1.9  &2.45 & 0.79 & 0.24& 5.3& 5.04  \\
$R_{\textrm{AP}}$[\%], LED intensity 2& 1.93 & 1.18 & 0.78 & 0.56  &0. 85& 7.31& 6.48  \\
$R_{\textrm{AP}}$[\%], LED intensity 3& 1.65 & 1.64 & 1.1&  0.78&  1. 09& 6.97&6.36 \\
$R_{\textrm{AP}}$[\%], average& 1.94  & 1.57  & 1.78& 0.71 & 0.73 & 6.53 & 5.96 \\
$R_{\textrm{AP}}$[\%], total & \multicolumn{4}{|l|}{$R^{\textrm{MCP}}_{\textrm{AP}}=5.66\pm 1.52$} & \multicolumn{3}{l}{$R^{\textrm{Dynode}}_{\textrm{AP}}=13.21\pm 2.28$}\\
\bottomrule
\end{tabular}
}
\label{tab:aptimeratio}
\end{table}
The stability of LED light intensity, afterpulse searching threshold and dark count of PMT are taken into account in the calculation of systematics. The uncertainties of the charge ratio calculation are estimated by considering both the experimental and statistical errors with conservative value used. 

A simplified afterpulse model can be built for JUNO 20-inch PMTs based on the afterpulse testing data with two assumptions: only dominant components of afterpulse in one time region are considered and the total charge ratio between afterpulses and primary pulse is calculated as the probability of occurrence for each afterpulse component. In this model, the afterpulse time distribution is described by Gaussian function, and its corresponding charge probability will be the coefficient as shown in Eq.~(\ref{eq:apd}) and Eq.~(\ref{eq:apm}):
\begin{equation}\label{eq:apd}
 \begin{split}
     P^{\textrm{MCP-PMT}}_{\textrm{AP}}=&\textrm{Gaus}(0.019,0.91,0.014)\\
     &+\textrm{Gaus}(0.016,3.134,0.061)\\
     &+\textrm{Gaus}(0.018,4.579,0.087)\\
     &+\textrm{Gaus}(0.0071,17.731,0.6),
\end{split}
\end{equation} 
\begin{equation}\label{eq:apm}
 \begin{split}
    P^{\textrm{dynode-PMT}}_{\textrm{AP}}=&\textrm{Gaus}(0.0073,1.067,0.132)\\
    &+\textrm{Gaus}(0.065,4.239,1.957)\\
    &+\textrm{Gaus}(0.06,15.081,2.791),
    \end{split}
\end{equation}
where $\textrm{Gaus}(A,\mu,\sigma)=A\cdot\textrm{exp}(-\frac{(t-\mu)^2}{2{\sigma}^2})$, $A$ is the occurrence probability of one afterpulse component when there is a primary pulse with charge of 1 p.e., $\mu$ is the average time of one group of afterpulses, and $\sigma$ is the standard deviation of the time distribution of one afterpulse component.
Considering all the confirmed afterpulse components,
the final afterpulse charge ratio of JUNO MCP-PMT (dynode-PMT) is 5.66\% (13.21\%), which is consistent with some previous measurement~\cite{TheDEAP:2017bxf,Wu:2019urv,Wu:2021}.

The afterpulse model defined in Eq.~(\ref{eq:apd}) and Eq.~(\ref{eq:apm}) are based on the testing data with primary pulse intensity 100 p.e. level. It is not clear whether it still works when the primary pulse is several p.e. level, since the dependency of afterpulses' charge and rate on primary charge intensity could be quite different at very low primary pulse intensity due to the complex structure and manufacture technology of these large size PMTs~\cite{coates1973origins,Wu:2019urv,ma2011time,zhao2016afterpulse,TheDEAP:2017bxf}.

\section{Conclusion and discussion}
Totally 150 MCP-PMTs and 11 dynode-PMTs have been tested to study the afterpulse timing and charge characters of JUNO 20-inch PMTs. For MCP-PMTs, four groups of afterpulse located at 0.9$~\mu$s, 3.1$~\mu$s, 4.6$~\mu$s, 17.74$~\mu$s after the primary pulse are confirmed, the averaged total afterpulse charge ratio is $5.6\pm1.5\%$. For dynode-PMTs, three groups of afterpulse located at 1.1$~\mu$s, 4.2$~\mu$s, 15.1$~\mu$s after the primary pulse are confirmed, the averaged total charge ratio is $13.2\pm2.3\%$. A simplified afterpulse model is constructed to describe the afterpulse time and charge distribution of JUNO 20-inch PMTs, which can benefit the simulation, calibration and events reconstruction of future JUNO run. 

\end{document}